# Aspects of Mechanical Engineering for Undulators

*Haimo Joehri*
Paul Scherrer Institut, Villigen, Switzerland

**Abstract**
This paper gives an overview about aspects of mechanical engineering of undulators. It is based mainly on two types that are used in the SwissFEL facility. The U15 Undulator is an example of an in-vacuum type and the UE38 is an APPLE-X type. It describes the frame, the adjustment of the magnets with flexible keepers and the adjustment of the whole device with eccentric movers.

## 1 Function and Types

### 1.1 Function

Undulators are used to generate synchrotron light for synchrotron light facilities. They guide the electron bunches from an electron accelerator to a slalom with magnets. For free electron lasers, several undulators are used in line. If this line is long enough, the electrons interact with the emitted beam. Inside a bunch, they arrange themselves into micro bunches. As a consequence of this, they are emitting coherent light.

### 1.2 Types

Undulators can be divided in different groups.

– Magnets inside the vacuum chamber or outside.

– Fixed gap or variable gap.

– Planar undulator or APPLE design (Fig. 1). APPLE stands for Advanced Planar Polarized Light Emitter. In undulators with APPLE design, the magnet rows can also be shifted along the beam. With that option, the electron bunches are moving in a helix. This allows the creation of polarized light.

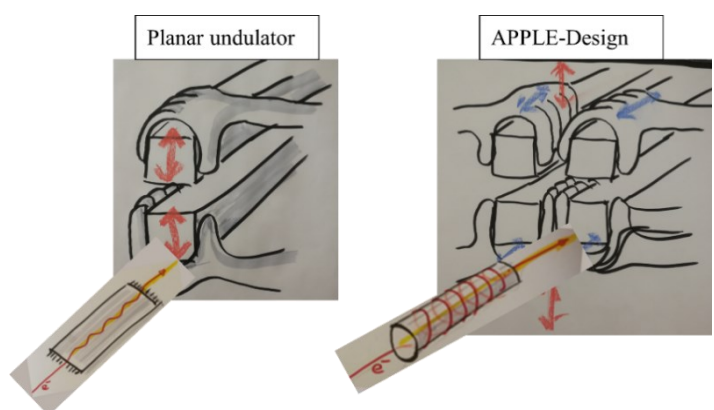

**Fig. 1:** Planar undulator and APPLE design

## 1.3 Examples

This report describes two examples of undulators from SwissFEL.

The U15 Undulator from the Aramis beamline of SwissFEL is a planar undulator with variable gap and the magnets inside the vacuum chamber (Fig. 2).

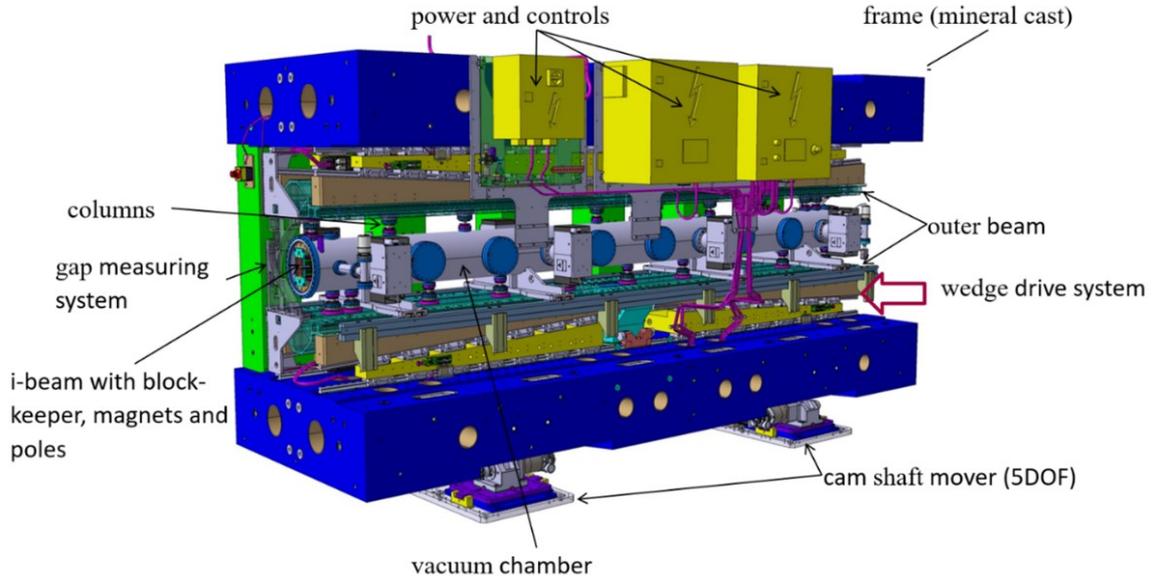

**Fig. 2:** Overview of U15 undulator (planar, in vacuum undulator)

The UE 38 from the Athos beamline of SwissFEL is an APPLE Undulator with variable gap and magnets outside the vacuum. The adjustment of the gap is arranged like an X (Fig. 3). Therefore, it is called APPLE-X Undulator.

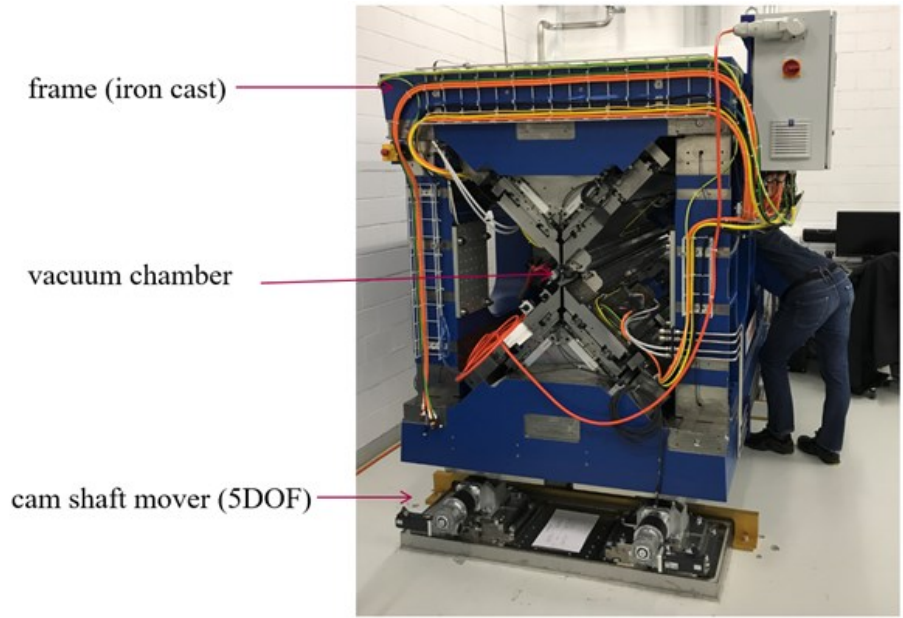

**Fig. 3:** Overview of UE38 undulator (APPLE-X design)

## 2 Frame

### 2.1 Forces

The forces of the magnets, that attracts each other can be several tons. In combination with the requirements of high precision, the challenges for the frame are quite high. In the past, most of the frames were designed in a C-Shape, because this gives the possibility to shift the vacuum chamber from the side into the frame during installation. To prevent the bending of the frame as consequence of the magnet forces, this design must be very heavy. For SwissFEL, we decided to go another way with a closed frame.

### 2.2 Example of U15 (Undulator for SwissFEL, Aramis Beamline)

For the U15 Undulator (Fig. 4), we chose mineral cast as material. The main reasons for this decision where:

– Cost savings: If you build a multitude of undulators, you have to invest in a casting mold, but the single frame is cost-efficient.

– Nonmagnetic: In some projects, this requirement is unclear, but with mineral cast, this is irrelevant.

– Possibility to integrate tubes in the frame.

– By agglutinate the blocks together, you receive a very stiff single part.

The disadvantages of mineral cast are mainly:

– cost of the casting mold, if only few parts are needed,

– young's modulus is lower than steel,

– thermal expansion is higher than steel.

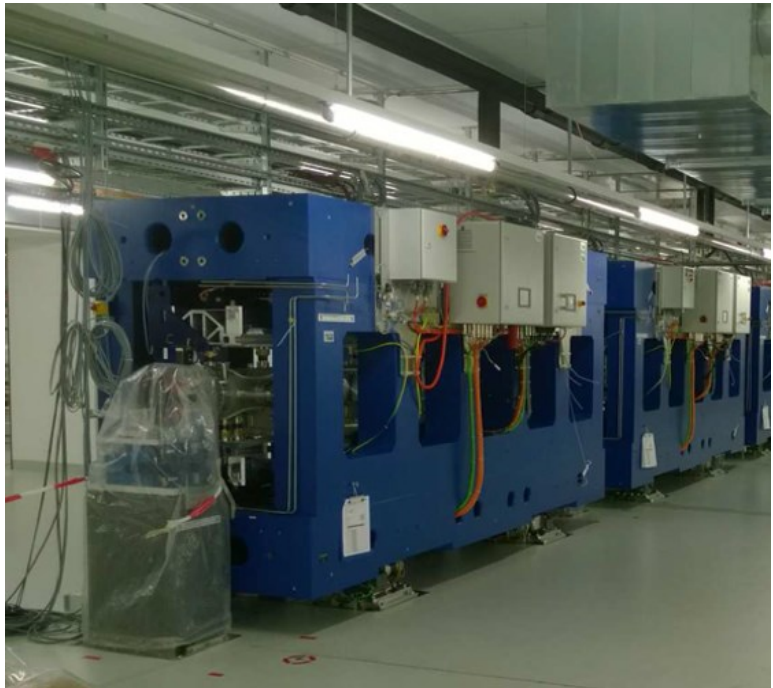

**Fig. 4:** U15 undulators in the SwissFEL tunnel

## 2.3 Example of APPLE-X Design (Undulator for SwissFEL, Athos-Beamline)

In the APPLE-X undulator, the arrangement of the guides is very sensitive to thermal expansion, because it is very much overdetermined (Fig. 5). Therefore, we decided to use cast iron for the frame, instead of mineral cast. All supporting elements are produced with the same material (Fig. 6). Instead of grinding, which is used for mineral cast, cast iron can be milled. This reduces the costs and gives more freedom in the design.

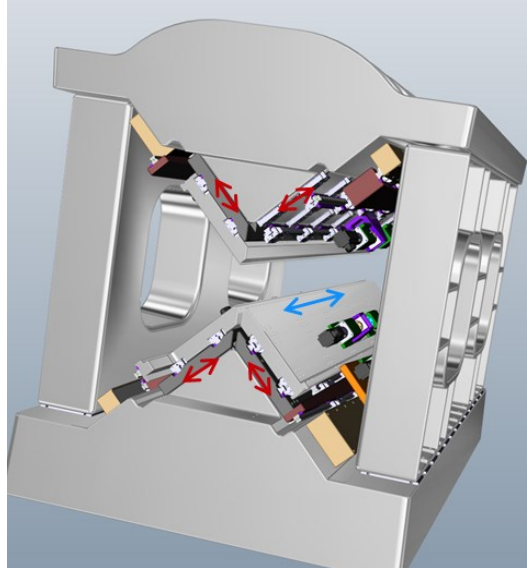

**Fig. 5:** Arrangement of UE38 Undulator

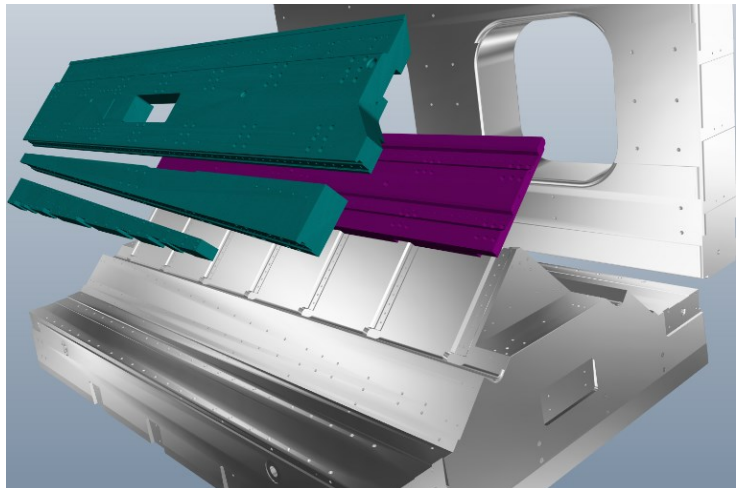

**Fig. 6:** Main parts of UE38 Undulators, all in cast iron

The design was developed by using tools for topology optimization. The first impression of these results looks very strange. But the main finding was, that a rip must be over each linear guide for the gap (Fig. 7). With that finding, the design was finally done with standard CAD-tools and optimized for casting. In general, tools for topology optimization are very useful, but the final design must be according to the manufacturing process.

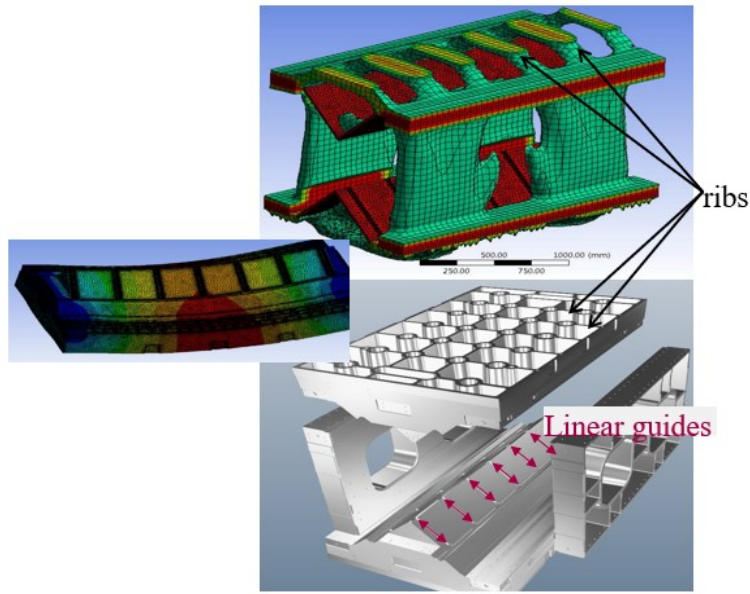

**Fig. 7:** Topology and simulation of UE38 Undulator frame

## 3 Keeper for the magnets

### 3.1 Basic challenges

An undulator consists of many magnets. For example, the U15 undulator has over 500 pairs of magnets. If you convert the magnetic tolerances in mechanical tolerances, the needed precision is below 10 microns. This tolerance must be fulfilled for each pair of magnets. In the past, the adjustment was done by shimming. This process can last several weeks for each undulator. For SwissFEL aramis beamline, we had to build 12 devices in a short time.

### 3.2 Idea of flexor keeper

A new idea was the development of a flexor keeper with the possibility to adjust each magnet by a screw. The upper part of the keeper is moved up and down by a wedge that is adjusted with a screw (Fig. 8). The material used for the keeper is aluminium. With that design, it is important to keep the flexor preloaded in each position, because the magnetic forces want to lift the magnet off the wedge. This gives a limitation of the range for adjustment which in this example is +/- 0.1 mm.

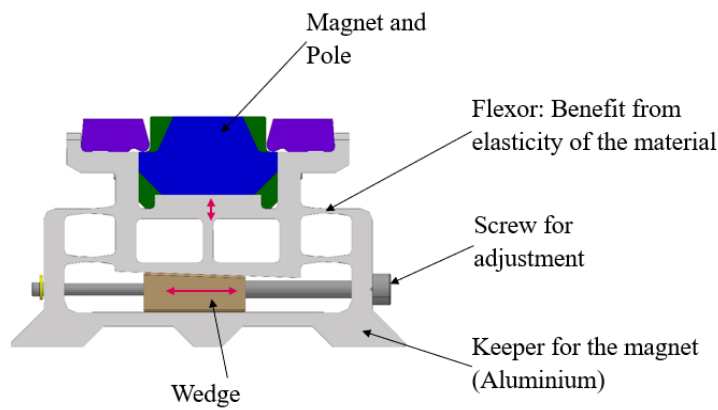

**Fig. 8:** Basic idea of flexor keeper

The first idea of fabrication was to produce these keepers by wire eroding, but during discussions with different engineering groups, we found optimized solutions. The next idea was to extrude it as a profile and cut this profile into pieces. But instead of cutting the complete profile, we finally cut only the upper part that must be flexible. As a final solution we got blocks with many keepers in one single part (Figs. 9 and 11).

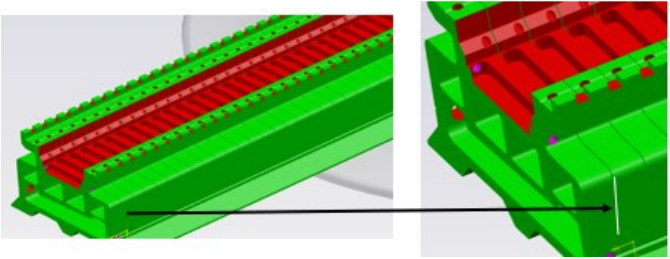

**Fig. 9:** Extruded profile with wire eroding of the flexible part

This design allowed us to develop a robot that can adjust each individual keeper in automated process. With the new keeper system of the U15 Undulator, we can adjust the 1000 Magnets of a 4 m long module with an automated robot system within 3 days.

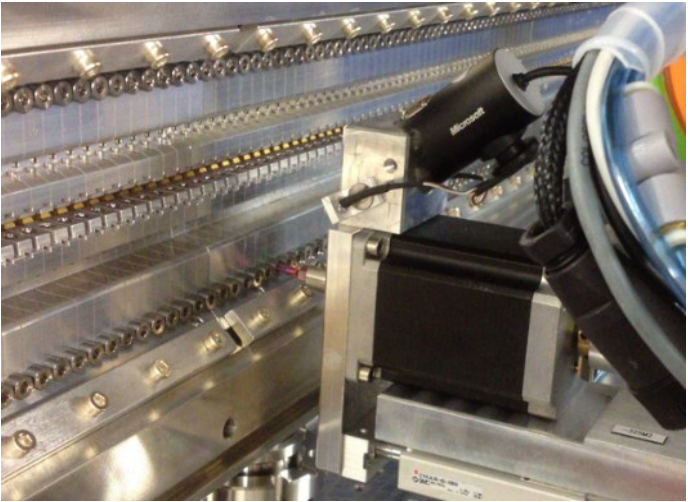

**Fig. 10:** Robot for adjustment

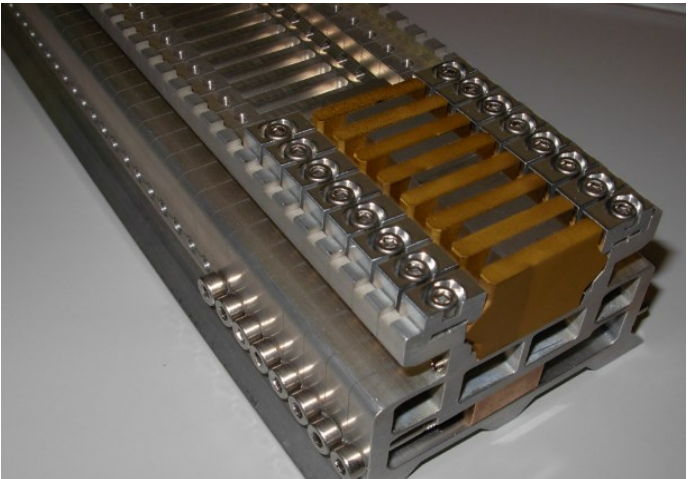

**Fig. 11:** Picture of a keeper with magnets

## 3.3 Further development for UE38

The UE38 Undulator has a period length of 38 mm instead of 15 mm in the U15 Undulator, this gives more space that allows some further developments. Mainly two features were changed:

- Preload of the flexor: Instead of using the flexor itself for the preloading against the magnetic forces, we added a separate spring (Fig. 12).
- Differential screw: For even more precise adjustment we changed the standard screw to a differential screw.

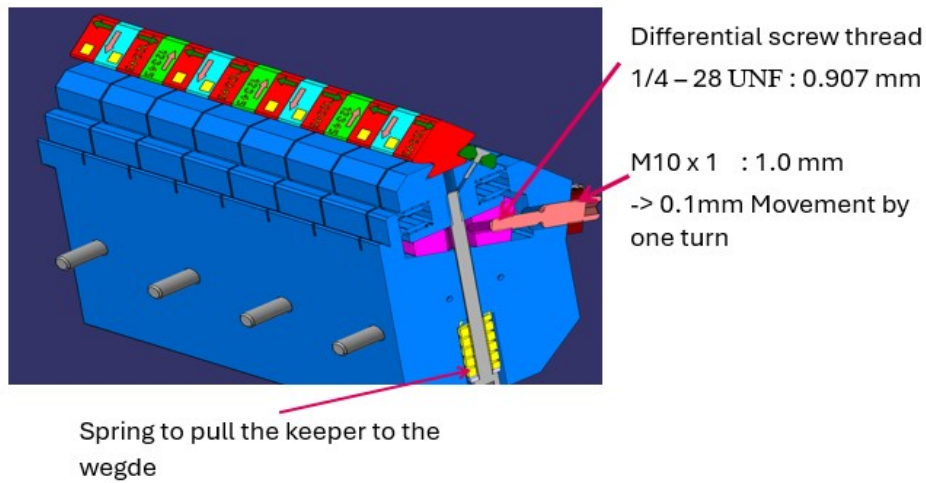

**Fig. 12:** Keeper for UE38 undulator

The differential screw consists of two threads, one thread of ¼-28 UNF with a slope of 0.907 mm and the other with standard M10 thread. With that system, by turning the screw 360°, the screw itself moves 1 mm, but the wedge only 0.1 mm (Fig. 13). Therefore, the resolution of one turn is 0.1 mm.

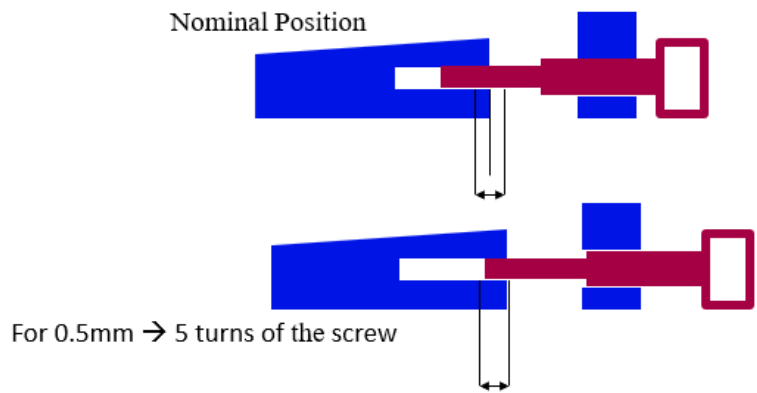

**Fig. 13:** Differential screw

The problem of that system is the starting point during assembly. If the position of the wedge in the starting point is shifted by one millimetre, the screw moves 10 mm to get into the correct starting point. This starting point is random because it depends on the starting angle of the threads. Due to that, a big traverse way for the screw is needed. To avoid that, the second thread is placed in a separate nut (Fig. 14), that is fixed after the correct placement of the wedge.

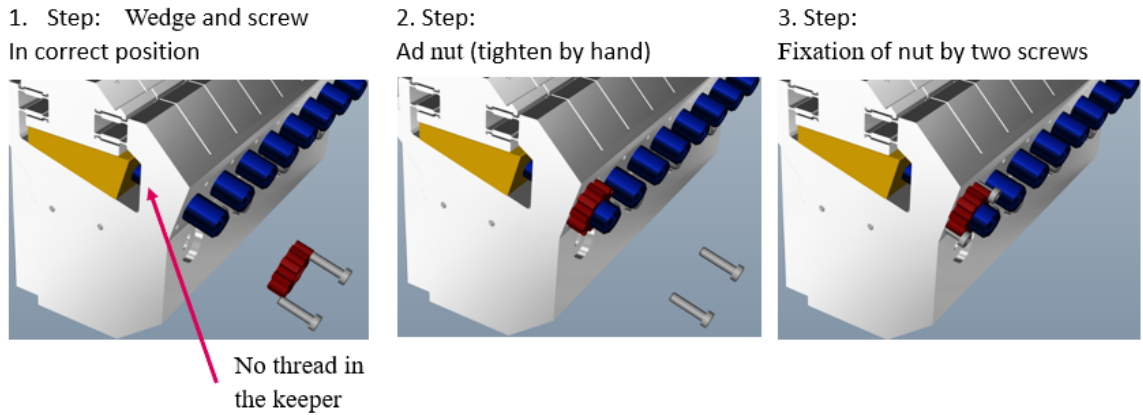

**Fig. 14:** Assembling process with differential screw

## 4 Columns

### 4.1 Basic Design

In in vacuum undulators, the keepers are mounted on a bar which holds the magnet keepers. This bar is connected to a mechanism that adjusts the gap by columns. All these columns must be adjustable with a resolution of one micrometre or even better. To reach this precision, backlash must be avoided. In our design, we had to add an additional thread (Figs. 15 and 16) to fix everything after adjustment [1].

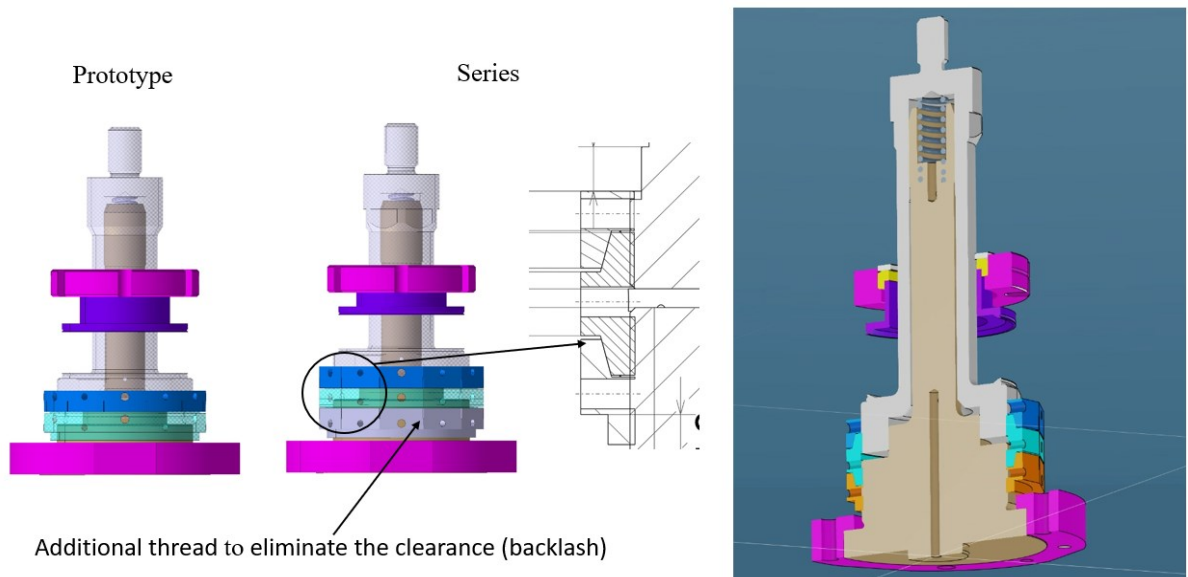

**Fig. 15:** Design of the column

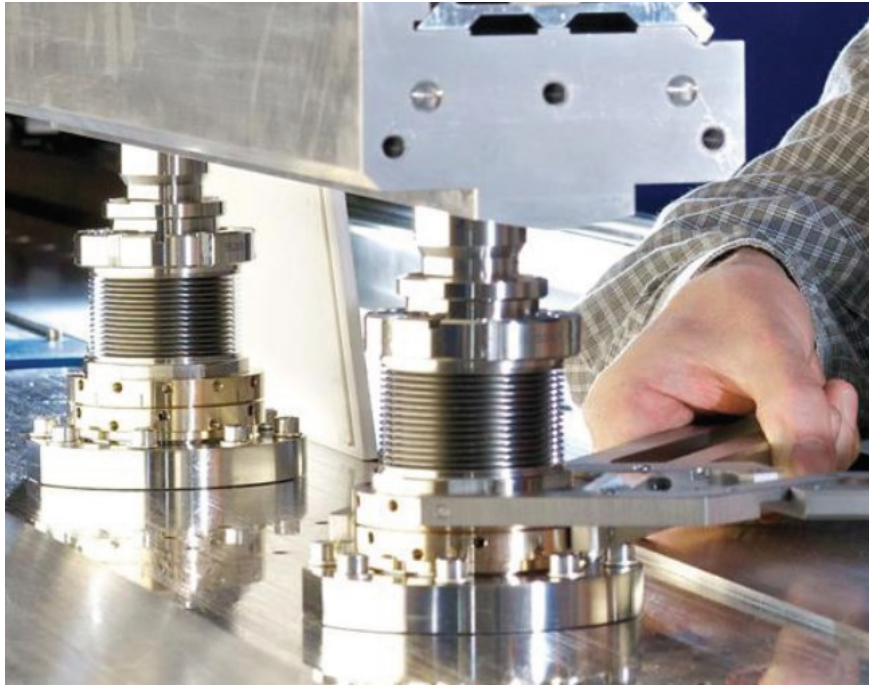

**Fig. 16:** Picture of the column adjustment

## 4.2 Arrangement

All magnets are adjusted for the nominal gap. This also compensate the magnet forces in the nominal position. As soon as the gap changes, also the forces changes. This effect will bend the bar that holds the magnet keepers (Fig. 17). Because this bar is supported only in single points by the columns, the resulting gap is not equal over the length.

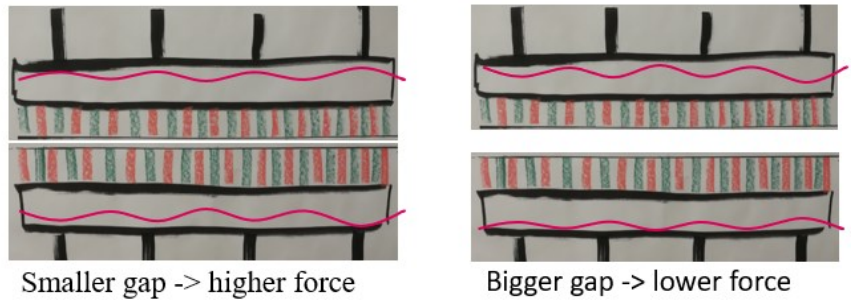

**Fig. 17:** Bending of supporting bar by smaller and bigger gap

To minimize that effect, that columns are arranged alternating (Fig 18). Even if the supporting bars are bending, the gap remains more constant [1]. The beam will not stay precisely in the middle of the magnetic field, but the field is pretty constant, so this effect can be neglected.

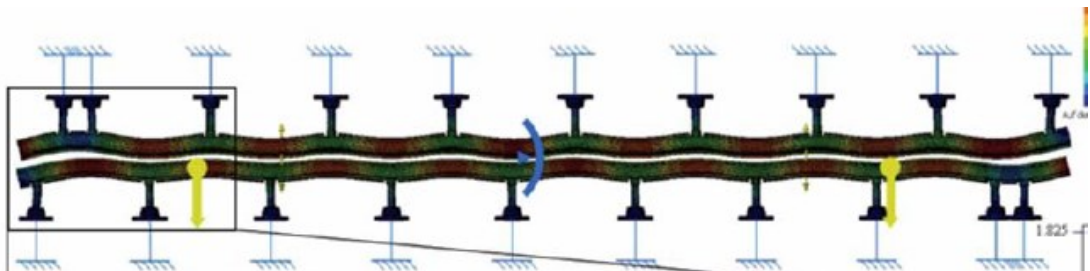

**Fig. 18:** Arrangement of the columns

## 5 Drives

For the drives, we used satellite roller screws (Fig 19). This has the advantage of a very small slope of 0.5 mm per turn and a small backlash. With such a small slope, the motor can be clutched directly to the spindle without any gear. This also helps to get a precise drive with very small backlash.

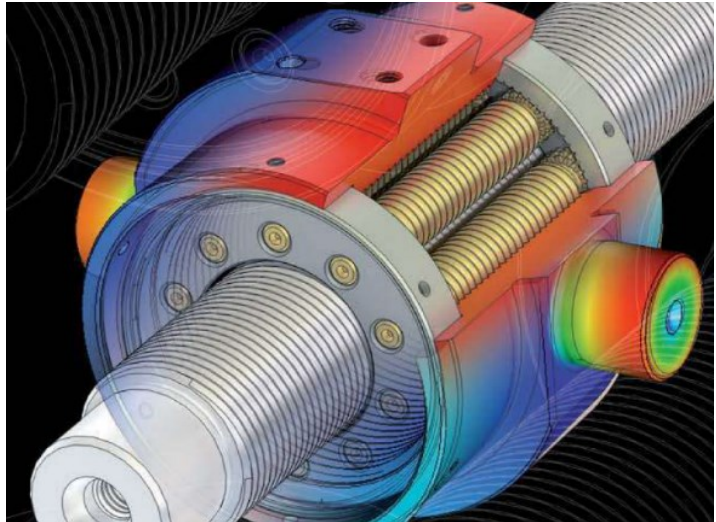

**Fig. 19:** Satellite roller screw

## 6 Mover

To adjust the whole device in the beam, we put the complete undulator onto movers which allows the adjustment in five degrees of freedom (Fig. 20). Along the beam axis, the position is less critical, therefore this axis is fixed and not adjustable. The movement works with eccentric bearings that are driven by servo motors. On one end of the device, two drives allow the adjustment vertical and to the side. On the other end, tree drives are used for vertical adjustment, adjustment to the side and also the turning around the beam axes (Figs. 21-24).

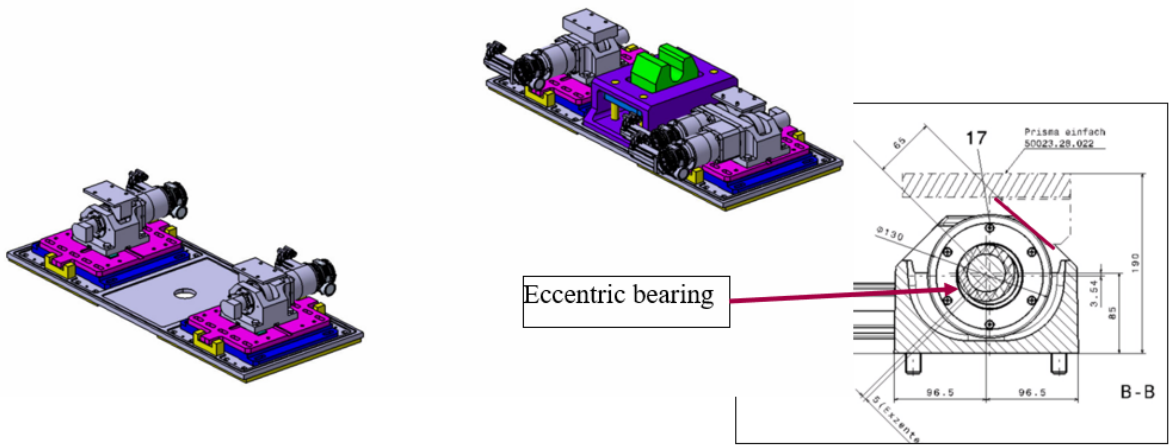

**Fig. 20:** Mover with 5 motors for 5 degrees of freedom

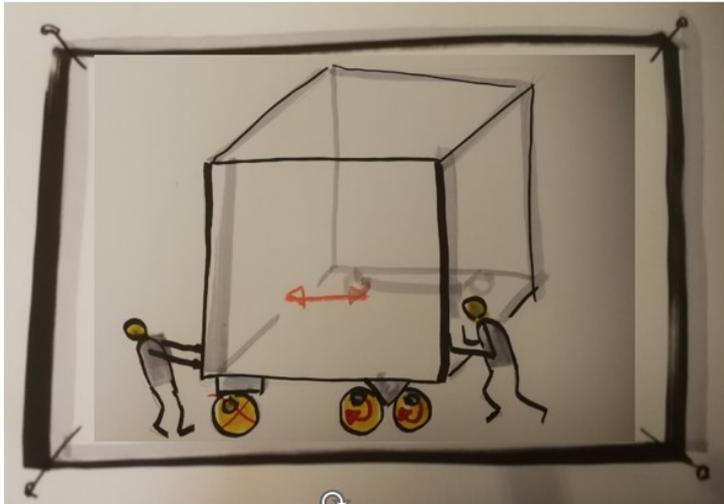

**Fig. 21:** Left-Right off frontside and turning around Z-axis

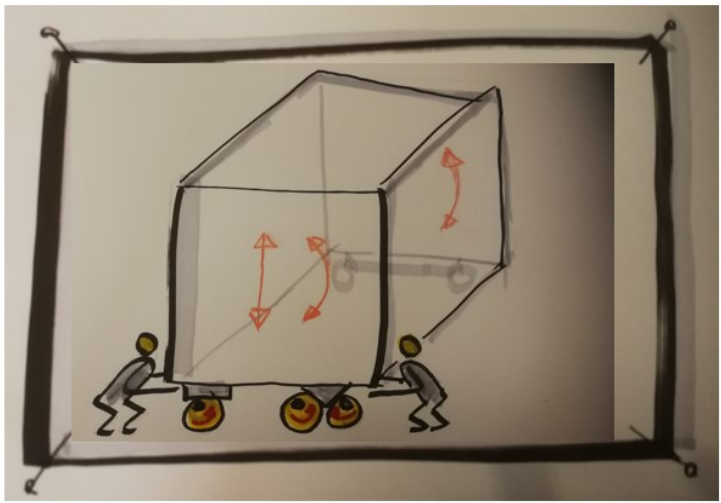

**Fig. 22:** Up-Down of frontside and turning around Y-axis

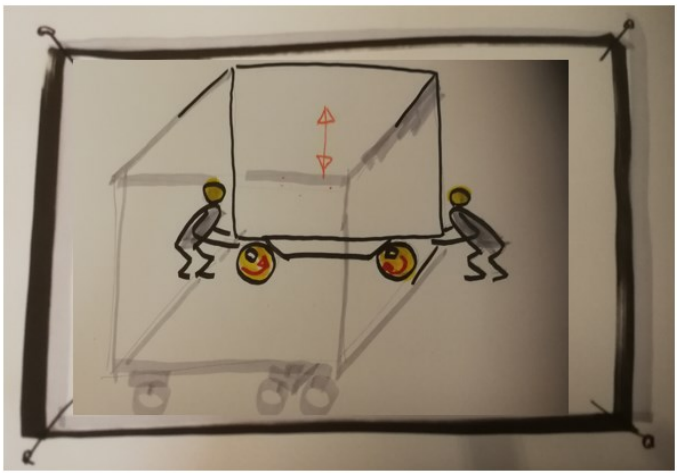

**Fig. 23:** Up-Down of backside and turning around Y-axis

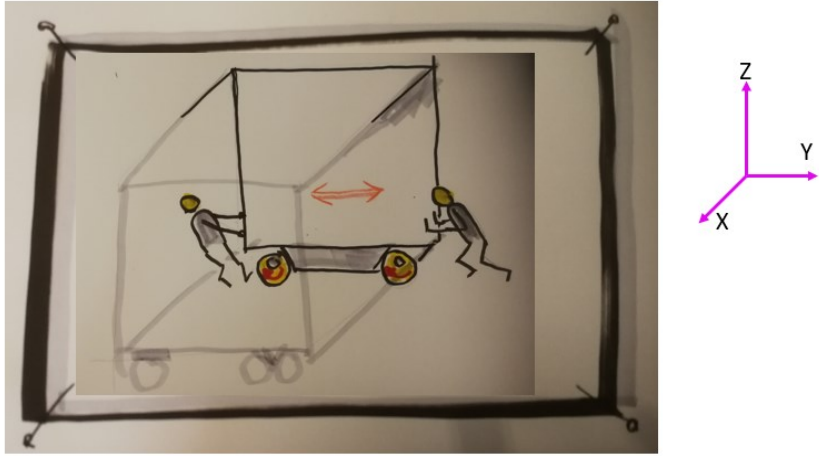

**Fig. 24:** Left-Right of backside and turning around Z-axis

## Acknowledgement

I want to thank the entire team of the mechanical engineering group and the colleges of the insertion device group.